# Task Frames

Burkhard D. Steinmacher-Burow

*Postfach 1163, 73241 Wernau, Germany*
*burow@ifh.de*
*http://www.tsia.org*

April 13, 2000

Forty years ago Dijkstra introduced the current conventional execution of routines. It places activation frames onto a stack. Each frame is the internal state of an executing routine. The resulting application execution is not easily helped by an external system.
This presentation proposes an alternative execution of routines. It places task frames onto the stack. A task frame is the call of a routine to be executed. The feasibility of the alternative execution is demonstrated by a crude implementation. As described elsewhere, an application which executes in terms of tasks can be provided by an external system with a transparent reliable, distributed, heterogeneous, adaptive, dynamic, real-time, parallel, secure or other execution. By extending the crude implementation, this presentation outlines a simple transparent parallel execution.

## 1 Introduction

This presentation proposes an alternative to the current conventional execution of routines. An application can use either or both executions. For example, the alternative execution thus introduces additional techniques for the implementation of a programming language.

The current conventional execution of routines generally requires an application to include the details of its reliable, distributed, heterogeneous, adaptive, dynamic, real-time, parallel, secure or other desired execution. Thus for example, a finance expert also needs to be a parallel execution expert in order to produce a finance application with a parallel execution.

In contrast, the alternative execution of routines allows an external system to provide an application with a transparent execution. A transparent execution is not visible to the application. This division of labour yields the usual benefits. Returning to the above example, a finance expert need not also be a parallel execution expert in order to produce a finance application with a parallel execution. Vice versa, the parallel execution expert need not also be an expert in finance nor in any other application domain in order to produce a system which provides an application with a transparent parallel execution.

Of course even with a transparent execution some details of the execution are part of the application. For example, such details may include realtime or reliability requirements or include constraints on the time or resource costs of the execution. Transparent are the execution details needed to meet such requirements and constraints. Another example is the demand by a transparent parallel execution that the application contains relatively obvious implicit parallelism. Transparent parallelism does not magically transform an inherently sequential application into one with implicit parallelism.

The alternative execution of routines is a result of the Task System and Item Architecture (TSIA), a model for transparent application execution. The name TSIA is used for both the model and for systems implementing the model. In many real-world projects, a TSIA provides a simple application with a transparent execution [Dividing]. TSIA is suitable for many applications, not just for the simple applications served to date. Various aspects of this suitability are shown elsewhere [Alternative][Dataflow][TSIA]. This presentation shows other aspects of this suitability.

The aspects of this presentation concern an alternative execution of routines. It places task frames onto a stack. A task frame is the call of a routine to be executed. By contrast, the current conventional execution of routines places activation frames onto a stack. An activation frame is the internal state of an executing routine.

TSIA is one of a succession of dataflow models. With TSIA, a simple and powerful dataflow model is achieved [Dataflow]. In the dataflow model, an executing application is a directed acyclic graph (dag). Each node of the graph is a task. Each arc of the graph is an item produced by one task and used by another. The execution of the application consists of executing the tasks of the graph. The execution of a task may change the graph and may use or produce the items of the task. The dataflow model thus is a form of graph reduction. For example, ALICE [ALICE] is an implementation of graph reduction with many similarities to the implementation of dataflow by Cilk-NOW and its precursors [Cilk-1][Cilk-NOW][PCM].

In the succession of dataflow models, that of Cilk-NOW and its precursors is the closest model to that of TSIA. Cilk-NOW provides a subset of the C programming language with a transparent reliable, adaptive and parallel execution. In Cilk-NOW and its precursors, an application executes in terms of tasks using a structure similar to a stack.

A TSIA using a stack of task frames, as in the alternative execution of routines, briefly has been introduced elsewhere [Dataflow]. Instead of a stack, other structures may be used for tasks and may be necessary for implementing some programming language features or execution features. Other presentations of TSIA generally refer to a task pool, thus not specifying the structure [Alternative][Dataflow][TSIA].

The current conventional execution of routines is briefly described in section 2. The alternative execution of routines is briefly described in section 3. In order to clearly present the alternative execution, section 4 briefly describes a programming language with a syntax close to the semantics of the alternative execution of routines. Section 5 shows a crude implementation of the programming language, thus demonstrating the feasibility of the alternative execution of routines. Section 6 shows that the current conventional execution of routines can be treated as a special case of the alternative execution. As shown in section 7, in the crude implementation of the alternative execution, the time overhead for calling a routine is about four times that in the current conventional execution. For many applications, such an additional overhead is small enough to be negligible. Similarly, as shown in section 8, iteration using tail recursion can have a time overhead within an order of magnitude as small as that of a loop. The current conventional execution of routines allows the amount of memory space occupied by local variables to depend on the values of one or more arguments of the routine. This also is allowed by the alternative execution, as demonstrated in section 9 using an array example. A simple transparent parallel execution is outlined in section 10. The alternative execution of routines is briefly compared in section 11 to some related work.

## 2  Activation Frames

The current conventional execution of routines is based on a stack of activation frames [Stack]. An activation frame is the internal state of an executing routine. When a routine is called, its activation frame is pushed onto the stack. When a routine exits, its activation frame is popped off the stack. At any given time, only the routine of the topmost frame is executing. A par-ent routine, suspended during the execution of its child, continues execution after the return of its child.

As an example, Figure 1b) shows snapshots of the stack during the execution of the routine `d` of the pseudocode in Figure 1a). If the state within an executing routine is ignored, then the snapshots show each distinguishable state in the execution. For simplicity, the arguments and bodies of the routines are elided.

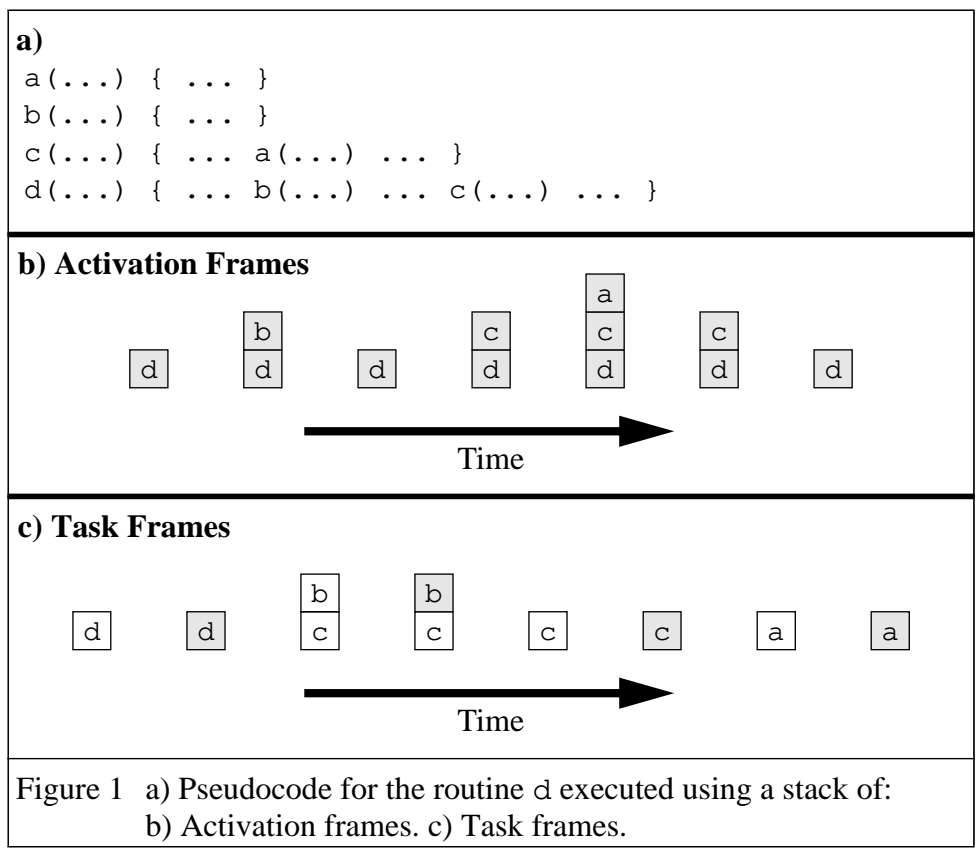

Figure 1  a) Pseudocode for the routine `d` executed using a stack of:
b) Activation frames. c) Task frames.

In general, an executing routine is supposed to efficiently produce its output and other external effects. An unstructured execution has more degrees of freedom than a structured execution and thus allows a more efficient execution. Thus a routine has an unstructured execution. In general, a routine need not have a structured execution since there is no external interest in the internal execution of a routine.

A structured execution moves from one structured state to another. A structured state clearly describes the items of the state and their dependen-cies. Since a structured execution clearly describes the execution, an external interest is better served by a structured execution than by an unstructured execution.

The internal state of an executing routine is given by an activation frame. It includes the state of local variables and the program counter. As described above, the state generally is unstructured. For example, an analysis might be required to determine the live variables. Dead variables are not items of the state. Thus an activation frame might not even clearly identify the items of the state, never mind their dependencies. In Figure 1b), the activation frames are shaded in order to illustrate their unstructured and hence opaque nature.

In the current conventional execution of routines, the state of an executing application is given by a stack of activation frames. Since each activation frame is unstructured, the current conventional execution of routines is an unstructured execution.

For an example of the unstructured execution, assume that the routine `d` in Figure 1a) has no code after calling the routine `c`. Then in the last four snapshots of Figure 1b), the entire frame `d` is dead and irrelevant to the state of the execution. In other words, the execution does not provide proper tail calls [RABBIT]. This lack of proper tail calling is one aspect of the unstructured nature of the current conventional execution of routines.

External interests in the application execution are not well served by an unstructured execution. For example, a stack of activation frames does not easily allow an external system to provide an application with a transparent reliable, distributed, heterogeneous, adaptive, dynamic, real-time, parallel, secure or other execution.

## 3  Task Frames

For the execution of routines, task frames are an alternative to activation frames. A task frame is the call of a routine to be executed.

Using a stack of task frames, routines execute as follows. The topmost task frame is popped from the stack and the call of the routine is executed. If the routine calls other routines, those task frames are pushed onto the stack. In other words, in its execution a task frame can replace itself by other task frames. Thus the new topmost frame is topmost among the replacement frames; otherwise it is the frame below the original topmost frame. The procedure repeats for the new topmost frame.

As an example, Figure 1c) shows snapshots of the stack during the execution of the routine d of the pseudocode in Figure 1a). If the state within an executing routine is ignored, then the snapshots show each distinguishable state in the execution.

In Figure 1c), initially on the stack is the task frame for the call to the routine d. Since it is the topmost frame, d executes and is popped from the stack. Since d calls b and c, their task frames are pushed onto the stack. Then b executes and its task frame is popped from the stack. In the execution of c, its task frame is replaced by one calling the routine a. Then a executes and its task frame is popped from the stack.

An application executed using task frames has the same result when executed using activation frames. In other words, the application definition is independent of its execution. For example, activation frames and task frames execute routines in depth-first order. Any input/output or other nonlocal effects of the routines thus have the same order in both executions. An application execution using either or both executions has the same result. Since the application execution maintains the same result under either execution, the execution of routines using task frames introduces additional techniques for the implementation of a programming language.

Once started, a task executes to completion, without waiting or suspending. In other words, the execution is nonblocking. This criterion defines a task.

Thus in an execution of routines using task frames, the execution of a child routine does not return to its parent. This is in contrast to an execution using activation frames. A task frame thus does not allow a parent to use the output of its child. This includes local items returned by the child as well as any nonlocal effects of the child.

For example, in routine c of Figure 1a), the output of the call to routine a is not used by the subsequent elided code. If this were not the case, the routine c easily could be automatically translated into conformance. The offending code is replaced by calls to routines containing that code. For the routine c, the translation might be:

```
h(...) { ... }
c(...) { ... a(...) h(...) }
```

Routine h contains the code dependent on the output of the call to a. Alternatively, instead of introducing the routine h, the original routine c could be executed using an activation frame. The ability to have task frames and activation frames on the stack is described in section 6.

As described in the previous section for an execution using activation frames, a routine has an unstructured execution. This also holds true when using task frames. Thus in Figure 1c) for task frames, as in Figure 1b) for activation frames, the frame of an executing routine is shaded in order to illustrate its unstructured and hence opaque nature.

In contrast to the execution of a routine, the call of a routine is a single structured state. A call, or equivalently its task frame, consists of the name of the routine and the arguments to the routine. In routine `d` of Figure 1a) for example, the call of routine `c` might be `c(7,v)`. Then in the execution of Figure 1c), the contents of the task frame `c` are the three items `c`, `7`, `v`. In Figure 1c), the task frames are unshaded in order to illustrate their structured and hence clear nature.

For example, unlike an activation frame, a task frame does not require analysis in order to determine the live items. The routine and its arguments generally are live. The declaration of the routine can state if they are not [Alternative].

In the alternative execution of routines, the state of an executing application is given by stack of task frames. Each task frame is structured. In addition, the dependencies between tasks are structured, as described in the next section. Thus the alternative execution of routines is a structured execution. For example, unlike the current conventional execution, the alternative execution provides proper tail calling.

A structured execution moves from one structured state to another. As illustrated in Figure 1c), between the execution of routines the stack of task frames is in a structured state. The state within an executing routine is not of external interest.

External interests in the application execution are well served by a structured execution. For example, a simple transparent parallel execution is outlined in section 10. Similarly, tasks allow an external system to provide an application with a transparent reliable, distributed, heterogeneous, adaptive, dynamic, real-time, secure or other execution [Dividing][TSIA].

## 4  A TSIA Language

The previous section introduced an application execution in terms of tasks. The execution is a result of the Task System and Item Architecture (TSIA), a model for transparent application execution. TSIA is described elsewhere [Dataflow][TSIA]. In this presentation, a task is a call of a routine. In par-ticular, this presentation focuses on the execution of routines using a stack of task frames.

As described in the previous section, for an execution using task frames, the state within an executing routine is not of external interest. Thus the implementation of a routine is completely open. For example, the routine could be implemented in a variety of programming languages. The use of task frames is not restricted to a particular programming language. Instead, task frames provide additional techniques to implement a programming language.

Following the real-world success of imperative languages and their extensions, this and other presentations of TSIA use an extended imperative language. A further description of the language and many application examples are available elsewhere [TSIA]. Since the syntax is close to the TSIA semantics, the language is called a TSIA language. By choice, the imperative part of the language is similar to the C programming language.

The imperative part of the language executes within a routine. The other part of the language is the task part and concerns calls to routines.

The TSIA language of this presentation is designed to help clearly present TSIA and to demonstrate its feasibility. A crude implementation is described in the next section. A TSIA language with a sophisticated implementation is described elsewhere [Cilk-1][Cilk-NOW][PCM].

For a routine written in a TSIA language, the execution proceeds as usual, except that calls to other routines yield the replacement tasks on the stack, as briefly described in the previous section. This alternative implementation of routines is known as delegation in other presentations of TSIA. Delegation also offers other benefits [Alternative][Dataflow][TSIA].

A task consists of items: ins, inouts and outs. An in is an item used by the task. An inout is an item modified by the task. An out is an item produced by the task. An inout behaves like an in and an out. An item can be of arbitrary size and complexity.

One of the ins is the instruction of the task and represents the actions executed by the task. An instruction can be as small as a single machine instruction or as large as a million-line program. In this presentation, an instruction is a routine. The syntax used is `routine(in,…; inout,…;out,…)`. The syntax is conventional, except that a semi-colon (`;`) separates the ins from the inouts and another semi-colon separates the inouts from the outs.

In order to demonstrate the TSIA language, Figure 2a) repeats the code of Figure 1a), but with the arguments and bodies of the routines. The rou-tines are used in Figure 2b), which shows snapshots of the stack during the execution of the application `d(;;q)`. The item `q` is assumed to be used by some subsequent task not included in the illustration here. The execution illustrated in Figure 2b) is like that in Figure 1c), except that Figure 2b) does not show snapshots of executing routines. Thus Figure 2b) only shows the structured states of the stack between the execution of routines. Figure 2b) includes comments showing the values of items in the snapshot. A crude implementation of this application execution using task frames is described in the next section.

An application execution using task frames is a directed acyclic graph (dag) [Dataflow]. Each node of the graph is a task. Each arc of the graph is an out of one task and an in of another. As given by the arcs, the dependencies between tasks are structured and explicit.

The execution of the application consists of executing the tasks of the graph. The execution of a task may change the graph and may evaluate the items of the task. An execution of routines using task frames thus is a form of graph reduction. This is illustrated in Figure 2c), using the execution of the application `d(;;q)`. In Figure 2c), each graph corresponds to a snapshot of the stack shown in Figure 2b). The application execution is a structured execution, moving from one structured state to another.

Instead of a graphical illustration like that in Figure 2c), this and other presentations of TSIA generally use a textual representation of a graph. For example, instead of using the second snapshot illustrated in Figure 2c), the presentations simply use the text `b(;w;) c(w;;q)`. A top-to-bottom and left-to-right order is used for the dependencies between tasks. A task producing an out precedes any task which uses that item as an in. For example, the graph `b(;w;) c(w;;q)` is not the same as the graph `c(w;;q) b(;w;)`. The order is familiar. For example, it is the order of an imperative programming language. The order is that of the TSIA language presented here.

The order is suitable for a sequential execution. As demonstrated in Figure 2b), the order of tasks on the stack is an example of such an order.

Since some tasks may be independent, the total order may not be unique. In other words, the dependencies between tasks may specify only a partial order. Independent tasks may be switched in the total order without affecting the results of the application. Independent tasks thus may execute in either order or in parallel.

In the graph of an application execution, an arc or item is unique, but its name is arbitrary. As long as all occurrences are changed, an arc or item

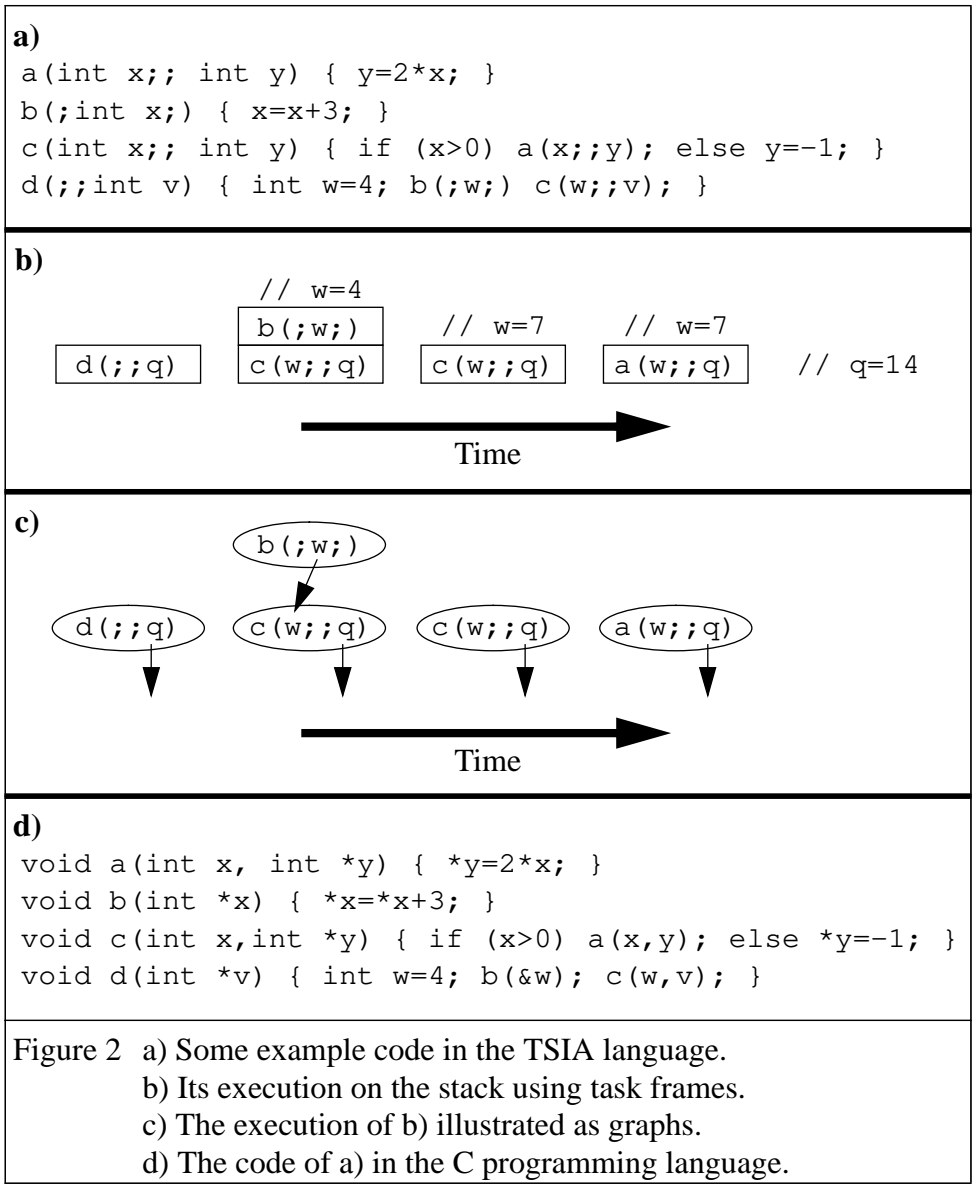

Figure 2  a) Some example code in the TSIA language.
b) Its execution on the stack using task frames.
c) The execution of b) illustrated as graphs.
d) The code of a) in the C programming language.

may be arbitrarily renamed. For example, the graph `b(;w;) c(w;;q)` of Figure 2b) or c) is equivalent to the graph `b(;v;) c(v;;q)`.

Instead of using task frames, the TSIA code of Figure 2a) could execute using activation frames. The latter execution would correspond to that of the C code shown in Figure 2d). Since the TSIA code of Figure 2a) does not use any language features not present in C, it only differs in syntax to the C code of Figure 2d). The execution is illustrated in Figure 1b). A crude implementation of this application execution using activation frames is described in section 6.

## 5 A Crude Implementation

This section shows a crude implementation of the TSIA language introduced in the previous section, thus demonstrating the feasibility of the alternative implementation of routines. The implementation uses C as an intermediate language. A compiler from TSIA code to the intermediate C code does not exist yet. The examples shown here are manually compiled.

In addition to the intermediate C code, an application also requires the code in Figure 3. The stack pointer `_sp` points to the topmost task frame and is initialized such that the stack is empty. The routine `_xtop` repeatedly executes the topmost task frame, until the stack is empty. The routine `_xtop` assumes that the first word of a task frame points to the routine to be called. The routine `_xtop` is one of several variations on a mechanism known as the UUO handler [ghc][RABBIT][sml2c].

```
#define _SSIZE 250000
int _stack[_SSIZE];            /* int for alignment */
char *_sp = (char*)(_stack+_SSIZE);

void _xtop(void) {
  while (_sp<(char *)(_stack+_SSIZE)) {
    (*((void(**)(void))_sp))();
  }
}
```

Figure 3  The stack and code to repeatedly execute the topmost task frame.

Figure 4 shows the intermediate C code resulting from compiling the TSIA code in Figure 2a). Figure 5 shows an application which calls the routine `d` of Figure 4. The code of Figures 3, 4, and 5 may be compiled and executed as a usual C application. As expected, the output is `14`.

In order to easily identify items introduced by the compilation, their names begin with an underscore. This also reduces the chance of their collision with names in the application.

In order to explain the compilation, attention initially is restricted to the routine `a(x;;y){y=2*x;}` of Figure 2a) and its intermediate C code in Figure 4. The compilation of the routine `a(x;;y)` yields the structure

```
extern char *_sp;    /* The stack pointer. */

/* a(int x;; int y) { y=2*x; } */
typedef struct {void (*_f)(void); int x;int *y;} _s_a;
void _f_a(void) {
  _s_a *_p = (_s_a *)_sp;
  *(_p->y) = 2*_p->x;
  _sp += sizeof(*_p);
}

/* b(;int x;) { x=x+3; } */
typedef struct {void (*_f)(void); int *x;} _s_b;
void _f_b(void) {
  _s_b *_p = (_s_b *)_sp;
  *(_p->x) = *(_p->x)+3;
  _sp += sizeof(*_p);
}

/* c(int x;; int y) { if (x>0) a(x;;y); else y=-1; }*/
typedef struct {void (*_f)(void); int x;int *y;} _s_c;
void _f_c(void) {
  _s_c *_p = (_s_c *)_sp;
  if ((_p->x)>0) {
    _s_a *_a1;
    /*_sp +=sizeof(*_p)-sizeof(_s_a);*/ /*Same size.*/
    _a1    = (_s_a *)_sp;
    _a1->_f = _f_a;
    /* _a1->x = _p->x; */ /* Already is true. */
    /* _a1->y = _p->y; */ /* Already is true. */
  }
  else {
    *(_p->y)=-1;
    _sp += sizeof(*_p);
  }
}
```

```
/* d(;;int v) { int w=4; b(;w;); c(w;;v); } */
typedef struct {void (*_f)(void); int *v;} _s_d;
void _f_d(void) {
  typedef struct { _s_b _b1; _s_c _c1;} _s_b_c;
  _s_b_c *_c;
  _sp += -sizeof(_s_b_c) + sizeof(_s_d);
  _c = (_s_b_c *)_sp;
  _c->_b1._f = _f_b;
  _c->_b1.x = &_c->_c1.x;

  _c->_c1._f = _f_c;
  _c->_c1.x = 4;
  /* _c->_c1.y Already is ok. */
}
```

Figure 4  Intermediate C code from compiling TSIA code in Figure 2a).

```
main() {
  int q;
  _s_d *s;
  _sp -= sizeof(_s_d);
  s = (_s_d *)_sp;
  s->_f = _f_d;
  s->v = &q;
  _xtop();
  printf("%d\n",q);
}
```

Figure 5  An application which calls the routine d of Figure 4.

{*_f,x,*y} of type _s_a, as well as the routine _f_a(). The structure describes a task frame for calling the routine. In the frame, _f points to the routine _f_a and is used to execute the frame. The routine _f_a has no usual arguments, since it receives its arguments x and y via the task frame. On entry to _f_a, the stack pointer _sp points to the frame of the call. _p gives structured access to the frame. Thus the original code y=2*x is compiled to *(_p->y)=2*_p->x. The in x is passed by value, while the out y is passed by reference. Since the execution of a frame removes it from the stack, the stack pointer is incremented by the size of the frame. The compilation of the routine `b` is very similar.

The compilation of the routine `c` introduces some aspects beyond those described above for the routine `a`. In particular, when `x>0` then the task `c(x;;y)` replaces itself by the task `a(x;;y)`. Since the two frames have the same size, the stack pointer does not change. The items `x` and `y` remain as is in the frame. Only `_f_c` is replaced by `_f_a` in the frame.

The compilation of `d` also introduces some aspects beyond those described above. In its execution, the routine `d` calls two routines. In other words, the task `d(;;v)` replaces itself by the tasks `b(;w;) c(w;;v)`. The structure `_c` of type `_s_b_c` is used for the two replacement frames. The stack pointer is moved to its new position. For the task `b(;w;)`, the pointers for the routine `_f_b` and for the argument `x` are written into its frame. For the task `c(w;;v)`, the pointer to the routine `_f_c` and the value `4` are written into its frame. In the replacement of the task `d(;;v)` by the tasks `b(;w;) c(w;;v)`, the item `v` remains as is on the stack.

Figure 5 shows an application which calls the routine `d` of Figure 4. First the stack pointer is decremented to make room for the frame, which is referred to using `s`. The pointers for the routine `_f_d` and for the argument `v` are written into the frame. The frame is executed by calling the routine `_xtop`, which returns when the stack is empty. In other words, when the call to `d` and all the resulting calls to routines have executed. As expected following the execution, the value of `q` is `14`.

Figure 2b) shows snapshots of the stack during the execution of the application `d(;;q)`. The crude implementation easily is instrumented to show the snapshots. In the loop of the routine `_xtop`, a call to the routine `_contents()` of Figure 6 is inserted before the execution of the topmost task.

```
/* Assume stack contains ints or pointers. */
void _contents(void) {
  int *p;
  for (p=(int*)_sp; p<_stack+_SSIZE; p++)
    printf("%8p:%8X==%10d\n",p,*p,*p);
  puts("End of snapshot of the stack.");
}
```

Figure 6  A routine to show the contents of the stack.

In the application of Figure 5, inserting the following code shows the addresses of interest.
```
    printf("_f_a==%p\n",_f_a);
    printf("_f_b==%p\n",_f_b);
    printf("_f_c==%p\n",_f_c);
    printf("_f_d==%p\n",_f_d);
    printf("&q ==%p\n", &q);
```
The output of the execution follows.
```
    _f_a==10c10
    _f_b==10c78
    _f_c==10ce8
    _f_d==10d88
    &q ==efffe838
      115450:    10D88==      69000
      115454:EFFFE838==-268441544
    End of snapshot of the stack.
      115444:    10C78==      68728
      115448:   115450==    1135696
      11544c:    10CE8==      68840
      115450:        4==          4
      115454:EFFFE838==-268441544
    End of snapshot of the stack.
      11544c:    10CE8==      68840
      115450:        7==          7
      115454:EFFFE838==-268441544
    End of snapshot of the stack.
      11544c:    10C10==      68624
      115450:        7==          7
      115454:EFFFE838==-268441544
    End of snapshot of the stack.
    14
```
The snapshots of the output and of Figure 2b) match exactly.

## 6  An Activation Frame is like a Task Frame

A suspended activation frame is similar to a task frame [Cilk-2]. The similarity is demonstrated in this section using a crude implementation of activation frames. The implementation is compatible with that of section 5 for task frames. The similarity allows an application execution to have a stack with activation frames and task frames. The current conventional execution of routines--using activation frames, thus can be treated as a special case of the alternative execution--using task frames.

As an example, the application of Figure 2a) may be compiled to have an execution using activation frames. The resulting intermediate C code for the routines c and d is shown in Figures 7 and 8, respectively. The call of a

```
extern char *_sp;    /* The stack pointer. */

/* c(int x;; int y) { if (x>0) a(x;;y); else y=-1; }*/
typedef struct {void (*_f)(void); int _entry;
                int x; int *y;} _s_c;
void _a_c(void) {
  _s_c *_p = (_s_c *)_sp;
  switch(_p->_entry) {
  case 0:;
    if ((_p->x)>0) {
      _s_a *_a1;
      _sp -= sizeof(_s_a);  _a1 = (_s_a *)_sp;
      _a1->_f = _f_a;  _a1->x = _p->x;  _a1->y = _p->y;
      _p->_entry=1; return;
  case 1:;
    }
    else *(_p->y)=-1;
  }
  _sp += sizeof(*_p);
}
void _f_c(void) {
  _s_c *_p = (_s_c *)_sp;
  _p->_f = _a_c;  _p->_entry = 0;
}
```

Figure 7  Routine c of Figure 2a) compiled to use an activation frame.

routine is a task frame, even when activation frames are used. In the execution of a task frame, its routine may convert the task frame to an activation frame. Such conversion is demonstrated by the compiled routines c and d.

Since the routine a does not call any routines, its execution using an activation frame is the same as that using a task frame. The same is true for

```
extern char *_sp;    /* The stack pointer. */

/* d(;;int v) { int w=4; b(;w;); c(w;;v); } */
typedef struct {void (*_f)(void); int _entry;
                int w; int *v;} _s_d;
void _a_d(void) {
  _s_d *_p = (_s_d *)_sp;
  switch(_p->_entry) {
  case 0:;
    _p->w = 4;
    { _s_b *_b1;
      _sp -= sizeof(_s_b);   _b1 = (_s_b *)_sp;
      _b1->_f = _f_b;   _b1->x = &_p->w;
      _p->_entry=1; return;
  case 1:;
    }
    { _s_c *_c1;
      _sp -= sizeof(_s_c);   _c1 = (_s_c *)_sp;
      _c1->_f = _f_c;   _c1->x = _p->w;   _c1->y = _p->v;
      _p->_entry=2; return;
  case 2:;
    }
  }
  _sp += sizeof(_s_d);
}
void _f_d(void) {
  _s_d *_p = (_s_d *)_sp;
  _p->_f = _a_d;   _p->_entry = 0;
}
```

Figure 8  Routine d of Figure 2a) compiled to use an activation frame.

the routine b. Thus the intermediate C code for the routines a and b in Figure 4, compiled in section 5 for an execution using task frames, also serves here for an execution using activation frames. Similarly used here is the application of Figure 5 which calls the routine d. The application is used in section 5 for an execution using task frames.

As for task frames, an execution using activation frames repeatedly executes the topmost frame. Thus the crude implementation of activation frames also uses the code of Figure 3.

Thus to demonstrate an execution using activation frames, the codes of Figures 3, 5, 7, 8 and the routines `a` and `b` of Figure 4 may be compiled and executed as a usual C application. As expected, the output is `14`.

As introduced in section 2, an activation frame is the internal state of an executing routine, including the state of local variables and the program counter.

For simplicity, the local variable `w` in the compiled routine `d` in Figure 8 is stored in the task frame. Space for `w` thus unnecessarily exists on the stack before the execution of the task. A better compilation might allocate space on the stack for local variables only when the task frame is executed and converts itself to an activation frame.

The C programming language, used for the output of the compiler, does not provide access to the program counter nor equivalently to alternate entry points for a routine. If it did, then `switch(_p->_entry)` and `case 0:` in the compiled routines `c` and `d` would not be required. Instead, `case 1:` and `case 2:` would be alternate entry points for the routine. For example, instead of `_p->_entry=1` in the compiled routine `c`, the code would be `_p->_f=_f_c_entry1`, where `_f_c_entry1` would be at the position of `case 1:`. At such an alternate entry point, variables like `_p` would have to be initialized or not subsequently used.

Since `_entry` is required in the crude implementation, it is initialized for the compiled routine `c` by the routine `_f_c`. The body of the routine `c` is placed into the routine `_a_c`. Thus even without alternate entry points, the call to `c` is a task frame.

As described in section 2, activation frames result in an unstructured execution. As described in section 3, task frames result in a structured execution. Of course, the results are relative. Since an activation frame is similar to a task frame, an execution using activation frames is not completely unstructured. For example, the structure of activation frames is sufficient to allow a debugger to travel the stack and to show the contents of each frame.

The execution of the above example application can demonstrate that task frames result in a more structured execution than that of activation frames. Figure 2b) shows snapshots of the stack for the execution using task frames. Figure 9 shows snapshots for the execution using activation frames. The execution illustrated in Figure 9 is like that in Figure 1b), except that Figure 9 includes all items of each frame. The execution using task frames illustrated in Figure 2b) is simpler and more structured than that illustrated in Figure 9 using activation frames. Nonetheless, Figure 9 illustrates that, as for task frames, also a stack of activation frames is a dag with an execution corresponding to graph reduction.

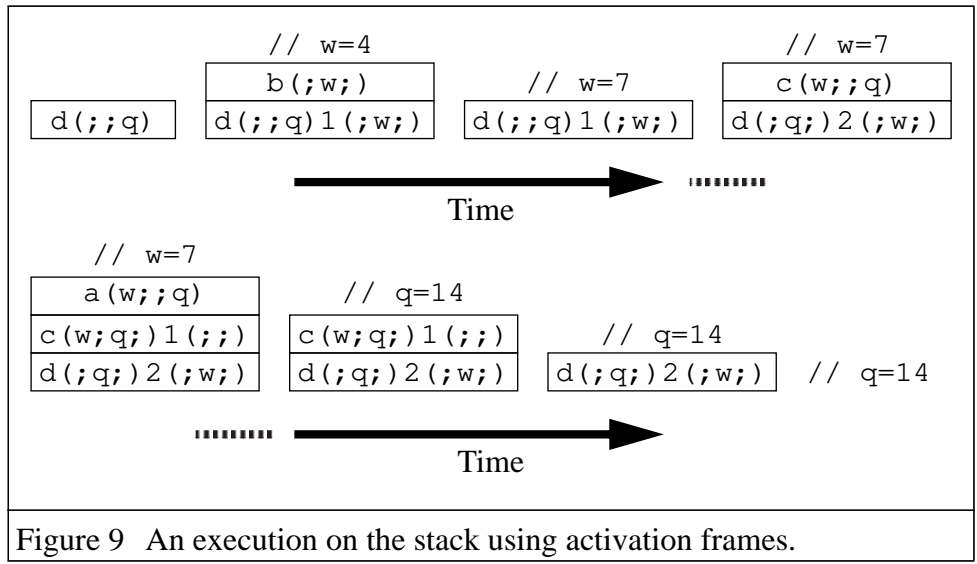

Figure 9  An execution on the stack using activation frames.

In Figure 9, the frame `d(;q;)2(;w;)` is to execute the code in routine `d` starting at the alternate entry point `2`. The local item `w` is used within the routine. Local items are declared in the parentheses following the entry point. Since the routine may read or write the item `q`, it is an inout of the frame. This is in contrast to the original frame `d(;;q)` where `q` is an out.

In the frame `d(;q;)2(;w;)` the items `q` and `w` are dead. They are not used in the execution of the routine `d` starting at the alternate entry point `2`. Thus after a live variable analysis, the frame could be written as `d(;;) 2(;;)`. Since it has no outs, the frame is dead and irrelevant to the remaining application execution.

As described in section 5, the crude implementation easily is instrumented to show snapshots of the stack during the execution. If a call to the routine `_contents` of Figure 6 is inserted to the routine `_xtop` of Figure 3, then the output of the above example application matches the snapshots shown in Figure 9.

As described above, the frame `d(;q;)2(;w;)` is dead and irrelevant to the remaining application execution. This is because the call `c(w;;v)` in `d` is a tail call, but an execution using activation frames does not provide proper tail calling. To prevent the useless frame `d(;q;)2(;w;)`, the rou-tine `d` could be compiled such that the frame `d(;;q)1(;w;)` executes as a task frame. In other words, the frame `d(;;q)1(;w;)` could replace itself by the frame `c(w;;q)`. In the compiled routine `d` of Figure 8, the call to `b(;w;)` remains unchanged. It continues to use an activation frame for `d`. In other words, the frame `d(;;q)` still replaces itself by the frames `b(;w;) d(;;q)1(;w;)`.

In a similar vein, an activation frame can place many task frames onto the stack. For example, `d` could be compiled such that the execution of the frame `d(;;q)` replaces itself on the stack by the frames `b(;w;) c(w;;q) d(;q;)3(;w;)`, where 3 is the suitable entry point.

The above examples demonstrate that a routine can be compiled to execute using task frames and activation frames. This is one possibility for an application execution to use task frames and activation frames. In a second possibility, each routine uses either a task frame or an activation frame, but the various routines of an application lead to a stack which contains both activation frames and task frames. In a third possibility, the stack consists only of task frames, but within the execution of a task a separate stack of activation frames is used. In other words, the use of a stack of activation frames can be encapsulated within a task. For example, a task can use a C library routine which executes using a stack of activation frames. The fourth possibility reverses the roles of the activation frames and task frames in the third possibility.

## 7 The Time Overhead for Calling a Routine

Section 5 showed a crude implementation of the alternative execution of routines. As described in this section, in the crude implementation the time overhead for calling a routine is about four times that in the current conventional execution. This demonstrates that the execution of routines using task frames can have a time overhead small enough to be practical for many parts of many applications.

The application used in this rough comparison is the doubly-recursive algorithm for computing the Fibonacci function. The algorithm is chosen for the comparison since it consists of many calls to routines and very little other work. The algorithm is a horribly inefficient computation of the Fibonacci function, but that is irrelevant here. Figure 10 shows the C code for a Fibonacci application using the conventional execution of routines.

Figure 11 also shows C code for a Fibonacci application using the current conventional execution of routines, but the application is written in the

```
int fib(int n) { return n<2 ? n : fib(n-1)+fib(n-2); }
main(){ int n = 36; printf("fib(%d)=%d\n",n,fib(n)); }
```

Figure 10 A conventional Fibonacci application.

task style. The routine `afib` calls the routine `add`, instead of directly using the output of its two `afib` children. The routine `add` is kept in a separate source file to prevent it from being inlined.

```
void add(int x, int y, int *z) { *z = x+y; }
void afib(int n,int *k) {
  if (n<2) *k = n;
  else
    {int h,j; afib(n-1,&h); afib(n-2,&j); add(h,j,k);}
}
main() { int n=36, k; afib(n,&k);
         printf("afib(%d,&k) yields k=%d\n",n,k); }
```

Figure 11 A Fibonacci application in the task style.

Figure 12 shows the intermediate C code resulting from compiling a Fibonacci application written in a TSIA language. The comments include the original TSIA code of the routines `tfib` and `tadd`. The application also requires the code in Figure 3 for the stack and the repeated execution of the topmost task frame. The code of Figures 3 and 12 may be compiled and executed as a usual C application. As expected, the output is `tfib(36;;k) yields k=14930352`.

Table 1 shows execution times for the above three Fibonacci applications. Computing environments vary widely. In order to show some of this variation, the execution times are shown for two processors, two compilers and two levels of optimization. The ratio `tfib/fib`, comparing the time of the crude implementation to that of the current conventional execution, is about four. A small part of this is due to the `add` routine required by the task style, as shown by the ratio `afib/fib` of about 1.4.

Compared to the above Fibonacci routines, most routines in other applications do more work. Other applications thus are not as sensitive to the time overhead for calling a routine. The above factor of four thus is a rough estimate of the time overhead for calling a routine in the crude implementation compared to that in the current conventional execution. The technique for the estimate also is used elsewhere [Cilk-1][Cilk-5].

```
extern char *_sp;    /* The stack pointer. */

/* tadd(int x, int y;; int z) { z=x+y; } */
typedef struct {void(*_f)(void); int x; int y;
                int *z;} _s_tadd;
void _f_tadd(void) {
  _s_tadd *_p = (_s_tadd *)_sp;
  *(_p->z) = _p->x + _p->y;
  _sp += sizeof(*_p);
}

/* tfib(int x;; int z)
   { if (x<2) z=x;
     else {tfib(x-1;;w); tfib(x-2;;v); tadd(w,v;;z);}
   }
 */
typedef struct{void(*_f)(void);int x;int *z;} _s_tfib;
void _f_tfib(void) {
  _s_tfib *_p = (_s_tfib *)_sp;
  if (_p->x<2) {*(_p->z) = _p->x; _sp += sizeof(*_p);}
  else {
    typedef struct {_s_tfib f1; _s_tfib f2;
                    _s_tadd a1;} _s_c;
    _s_c *_c;
    _sp += sizeof(*_p) - sizeof(_s_c);
    _c = (_s_c *)_sp;
    _c->f1._f = _f_tfib;
    _c->f1.x  = _p->x-1;
    _c->f1.z  = &_c->a1.x;

    _c->f2._f = _f_tfib;
    _c->f2.x  = _p->x-2;
    _c->f2.z  = &_c->a1.y;

    _c->a1._f = _f_tadd;
    /* _c->a1.z = _p->z; */ /* Already is true. */
  }
}
```

```
main() {
  int n=36, k;
  _s_tfib *b;
  _sp -= sizeof(_s_tfib);
  b = (_s_tfib *)_sp;
  b->_f = _f_tfib;
  b->x  = n;
  b->z = &k;
  _xtop();
  printf("tfib(%d;;k) yields k=%d\n",n,k);
}
```

Figure 12 Intermediate C code from compiling a Fibonacci application written in a TSIA language.

| Processor | Compiler | Seconds for Execution | | | afib/ fib | tfib/ afib | tfib/ fib |
|---|---|---|---|---|---|---|---|
| | | fib | afib | tfib | | | |
| sparc | cc | 5.5 | 5.0 | 13.9 | 0.9 | 2.8 | 2.5 |
| | cc -fast | 1.9 | 2.3 | 7.6 | 1.2 | 3.3 | 4.0 |
| | gcc | 3.4 | 5.0 | 13.4 | 1.5 | 2.7 | 3.9 |
| | gcc -O2 | 1.9 | 2.6 | 7.9 | 1.4 | 3.0 | 4.2 |
| i686 | gcc | 2.5 | 3.9 | 9.4 | 1.6 | 2.4 | 3.8 |
| | gcc -O2 | 2.8 | 4.3 | 6.0 | 1.5 | 1.4 | 2.1 |

Table 1: Execution times and ratios for three Fibonacci applications.

The factor four time overhead introduced by the crude implementation of the alternative execution of routines is comparable to that introduced by the various versions of Cilk. For many applications, Cilk has shown that such an additional overhead for calling a routine is small enough to be negligible [Cilk-1][Cilk-5].

## 8 Iteration using Tail Recursion

As long promoted and often practiced in the functional programming community, iteration can be performed using tail recursion. The currently more popular alternative uses a loop for iteration. That iteration efficiently can use tail recursion easily is demonstrated for an execution of routines using task frames.

An example of iteration using a loop is in the C routine `lsum` of Figure 13. The result of the iteration is `a0+i+(i+1)+..+n`. The result is more efficiently computed as `a0+(n+i)*(n-i+1)/2`. The example thus is a horribly inefficient computation, but that is irrelevant here.

```
void lsum(int i, int n, int a0, int *a) {
  for ( ; i<=n; i++) { a0 += i; }
  *a=a0;
}
```

Figure 13 Iteration using a loop.

Instead of using a loop, the above iteration can use tail recursion, as demonstrated by the C routine `csum` in Figure 14. Because it uses the current conventional execution of routines, the routine `csum` has a space-inefficient execution. Each iteration places an additional activation frame onto the stack. The space required thus is proportional to the number of iterations. In other words, there is no proper tail calling. In contrast, a loop executes within a single activation frame and thus requires only a constant amount of space, regardless of the number of iterations.

```
void csum(int i, int n, int a0, int *a)
{ if (i<=n) csum(i+1,n,a0+i,a); else *a = a0; }
```

Figure 14 Iteration using tail recursion with activation frames.

The C routine `csum` of Figure 14 differs only in name and syntax from the TSIA routine `tsum` in a comment of Figure 15. However their executions are very different. The iteration using tail recursion in `tsum` executes like a loop. In other words, there is proper tail calling. The execution uses a single task frame and thus requires only a constant amount of space, regardless of the number of iterations. This is evident in the intermediate C code of Figure 15 resulting from compiling the TSIA routine `tsum`. The intermediate C code relies on the code in Figure 3 for the stack and the repeated execution of the topmost task frame. For `tsum`, tail recursion corresponds to the iteration of repeatedly executing the topmost task frame. In other words, for an execution of routines using task frames, tail recursion is a loop.

```
extern char *_sp;    /* The stack pointer. */

/* tsum(int i, int n, int a0;; int a)
   { if (i<=n) tsum(i+1,n,a0+i;;a); else a=a0; } */
typedef struct {void (*_f)(void); int i; int n;
                int a0; int *a;} _s_tsum;
void _f_tsum(void) {
  _s_tsum *_p = (_s_tsum *)_sp;
  if (_p->i <= _p->n) {
    _p->a0 += _p->i;   /* Tail recursive,        */
    _p->i++;           /*  so just update arg.s. */
  }
  else { *(_p->a) = _p->a0;
         _sp += sizeof(_s_tsum); }
}

void sum(int i, int n, int a0, int *a)
{ _s_tsum *b;
  _sp -= sizeof(_s_tsum);
  b = (_s_tsum *)_sp;
  b->_f = _f_tsum;  b->i = i;   b->n = n;
  b->a0 = a0;       b->a = a;
  _xtop();
}
```

Figure 15 Intermediate C code from compiling `tsum`,
which performs iteration using tail recursion with a task frame.

Figure 16 shows an application using the routine `lsum` of Figure 13 to calculate `1+2+..+n`. Alternatively, the application can use the routine `csum` of Figure 14 or `tsum` of Figure 15. Convenient use of `tsum` is given by the routine `sum` in Figure 15.

Table 2 shows execution times for the above three applications for a small variety of computing environments. Compared to those of the above applications, most iterations in other applications do more work. Other applications thus are not as sensitive to the time overhead of an iteration. Thus the ratios of execution times in Table 2 are a rough estimate of the relative time overheads for the three implementations of iteration presented here.

```
main() {
  int j,e,a,n=64000;
  e = n%2? n*((n+1)/2):(n/2)*(n+1); /*avoid overflow*/
  for (j=0; j< 1000; j++) {
    lsum(1,n,0,&a);
    if ( a != e)
      printf("ERROR: 1+2+..+%d==%d NOT %d\n",n,e,a);
  }
}
```

Figure 16 An application using the routine `lsum` of Figure 13.

| Processor | Compiler | Seconds for Execution | | | csum/ lsum | tsum/ csum | tsum/ lsum |
|---|---|---|---|---|---|---|---|
| | | lsum | csum | tsum | | | |
| sparc | cc | 2.6 | 54 | 10.7 | 20 | 0.2 | 4.1 |
| | cc -fast | 0.30 | 0.30 | 4.9 | 1.0 | 16 | 16 |
| | gcc | 2.8 | 57 | 9.7 | 20 | 0.2 | 3.5 |
| | gcc -O2 | 0.30 | 54 | 4.7 | 180 | 0.1 | 16 |
| i686 | gcc | 1.0 | 14.8 | 4.0 | 15 | 0.3 | 4.0 |
| | gcc -O2 | 0.30 | 18.0 | 1.9 | 60 | 0.1 | 6.3 |

Table 2: Execution times and ratios for three implementations of iteration.

In Table 2, the large values for csum/lsum show a large time-inefficiency for iteration using tail recursion with activation frames. Presumably, the single activation frame of a loop uses registers and at most the cache, while the many activation frames of recursion do not even fit in the cache.

In Table 2, the values for tsum/lsum show that iteration using tail recursion with a task frame can have a time overhead within an order of magnitude as small as that of a loop. This is small enough to be practical for many iterations in many applications. In a real-time application, for example, the fine granularity required for tasks may not allow a loop to perform all its iterations within a single task [RTU]. Instead, the fine granularity can be met by a loop performing some of the iterations inside a tail recursive routine.

# 9 Arrays

For the execution of a routine using an activation frame, the amount of memory space occupied by local variables can depend on the values of one or more arguments of the routine [Stack]. For example, one of the arguments can be the length of a local array.

Likewise, for the execution of a routine using a task frame, the amount of memory space occupied by local items can depend on the values of one or more ins of the routine. The above array example is demonstrated using the routine `esum` of Figure 17. In addition to the intermediate C code resulting from compiling `esum`, Figure 17 shows the original TSIA routine in a comment.

The routine `esum` uses an array `v` of length `len=n-i+1` to compute `a0+i+(i+1)+..+n`. Different implementations of the same computation are presented in the previous section. The code of Figures 3, 16, and 17 may be compiled and executed as a usual C application. The application also requires a routine `sum` for `esum`, like that of Figure 15 for `tsum`.

The routine `esum` uses the routines `vseq` and `vsum`, also in Figure 17. The TSIA language of this presentation allows unambiguous shorthand. For example, in the call `vseq(len,i;;v)` in the TSIA routine `esum`, the out `v` implicitly is declared to be an array of length `len` due to the declaration `vseq(n,m;;a[n])` for the routine `vseq`. A longhand version of the call could be `vseq(len,i;;v[len])`.

The snapshots of the stack shown in Figure 18 illustrate an execution of the routine `esum`. Originally the task `esum(i,n,a0;;a)` is on the stack. In its execution, the task replaces itself on the stack by the contents shown in the second snapshot of Figure 18. Similar to previous examples of task execution, the new stack contents include task frames for the calls to the routines `vseq` and `vsum`. The contents also include a call to the internal routine `_skip` and the array `v`. The routine `_skip(n;;)` skips over the subsequent `n` bytes on the stack. The array `v` is part of the `_skip` task frame. After the execution of the tasks `vseq` and `vsum`, the array `v` no longer is required, so the execution of `_skip` removes the array from the stack. Even though it does not provide a proper tail call, this implementation of arrays is sufficient for many purposes.

In the intermediate C code for `esum` in Figure 17, the local variables `_i` and `_a` are introduced because the items `_p->i` and `_p->a` of the original task frame may be lost when the frame is replaced by the new stack contents.

```
extern char *_sp;    /* The stack pointer. */

/* _skip(n;;) Internal: Skip next n bytes on stack. */
typedef struct {void (*_f)(void); int n;} _s__skip;
void _f__skip(void) {
  _s__skip *_p = (_s__skip *)_sp;
  _sp += sizeof(*_p) + _p->n;
}

/* Set a[0:n-1] = m,m+1,...,m+n-1 */
/* vseq(int n, int m;; int a[n])
   { if (n>0) { a[0]=m; vseq(n-1,m+1;;a[1]); } } */
typedef struct {void (*_f)(void); int n; int m;
                int *a;} _s_vseq;
void _f_vseq(void) {
  _s_vseq *_p = (_s_vseq *)_sp;
  if (_p->n > 0) {
    _p->a[0] = _p->m;  /* Tail recursive,      */
    _p->n--;           /*  so just update arg.s. */
    _p->m++;
    _p->a++;
  } else _sp += sizeof(*_p);
}

/* z += a[0]+a[1]+..+a[n-1] */
/* vsum(int n, int a[n]; int z; )
   { if (n>0) { z+=a[0]; vsum(n-1,a[1];z;); } } */
typedef struct {void (*_f)(void); int n; int *a;
                int *z;} _s_vsum;
void _f_vsum(void) {
  _s_vsum *_p = (_s_vsum *)_sp;
  if (_p->n > 0) {
    *(_p->z) += _p->a[0];  /* Tail recursive,      */
    _p->n--;               /*  so just update arg.s. */
    _p->a++;
  } else _sp += sizeof(*_p);
}
```

```
/* esum(int i, int n, int a0;; int a)
   { int len=n-i+1; a=a0;
     if (len>0) { vseq(len,i;;v); vsum(len,v;a;); }
   } */
typedef struct {void (*_f)(void); int i; int n;
                int a0; int *a;} _s_esum;
void _f_esum(void) {
  _s_esum *_p = (_s_esum *)_sp;
  int len = _p->n-_p->i+1;
  *(_p->a) = _p->a0;
  _sp += sizeof(*_p);
  if (len>0) {
    int *v, _i=_p->i, *_a=_p->a;
    _s__skip *_p_skip; _s_vseq *_pvseq;
    _s_vsum *_pvsum;

    _sp -= len*sizeof(int);        v =     (int *)_sp;

    _sp -= sizeof(_s__skip); _p_skip =(_s__skip *)_sp;
    _p_skip->_f = _f__skip;
    _p_skip->n  = len*sizeof(int);

    _sp -= sizeof(_s_vsum);   _pvsum = (_s_vsum *)_sp;
    _pvsum->_f = _f_vsum;
    _pvsum->n  = len;
    _pvsum->a  = v;
    _pvsum->z  = _a;

    _sp -= sizeof(_s_vseq);   _pvseq = (_s_vseq *)_sp;
    _pvseq->_f = _f_vseq;
    _pvseq->n  = len;
    _pvseq->m  = _i;
    _pvseq->a  = v;
  }
}
```

Figure 17 Intermediate C code from compiling `vseq`, `vsum`, `esum`.

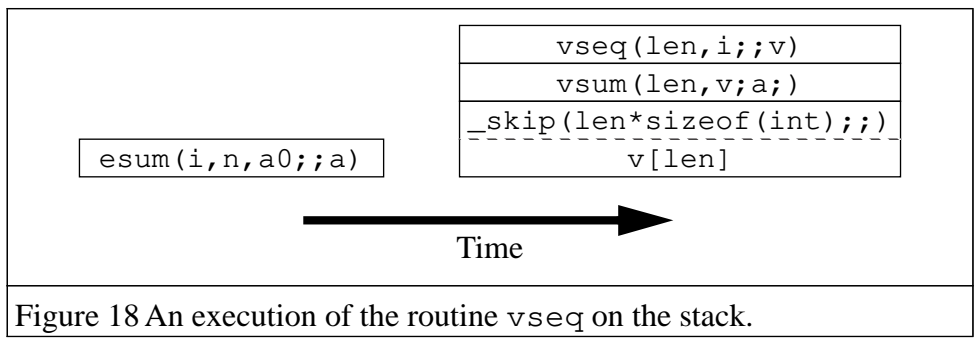

Figure 18 An execution of the routine `vseq` on the stack.

As introduced in section 5, in the implementation of this presentation, scalar ins are passed by value. Alternatively, scalar ins could be passed by reference, similar to the array of the above example.

## 10 A Transparent Parallel Execution

This section outlines TPARX (pronounced "tea park"), a simple implementation of a transparent parallel execution.

TPARX uses the TSIA language of this presentation and its crude implementation. TPARX requires three small changes to the compilation of TSIA code to the intermediate C code.

First, TPARX must be able to travel the stack, even though each frame has a different length. For example, a task frame thus might include its length at a fixed offset in the task frame.

Second, TPARX allows an application execution to use multiple stacks, each with its own stack pointer. For example, the routine of a task frame thus might have a pointer to the stack pointer as its argument. A corresponding change also would be made to the routine `_xtop` of Figure 3.

Third, TPARX requires a task frame to include a `_ready` flag at a fixed offset in the task frame. The values `_ready=1` and `_ready=0` indicate whether or not the task frame is ready to execute, even if the frame is not the topmost frame on the stack. As before, the topmost frame is ready to execute, regardless of its `_ready` flag. The value `_ready=1` thus identifies frames which may execute in parallel with each other and with the topmost frames. The value `_ready=0` ensures that the dependencies between tasks are met in the execution.

The `_ready` flag is set when the task frame is written to the stack. A task frame is written to the stack in the execution of the task's parent. Thus a task's `_ready` flag is set in the execution of the task's parent. For each of its children, a parent thus has an expression which yields the value of the
`_ready` flag. Often the expression is a compile-time constant.

The above three changes are illustrated using the routine `dcvsum` of Figure 19. The routine is a divide-and-conquer (DC) version of the routine

```
/* z = a[0]+a[1]+..+a[n-1] */
dcvsum(int n, int a[n];; int z )
{ if (n==1) z=a[0];
  else { int k=n/2;
         dcvsum(k  ,a    ;;rx);
         dcvsum(n-k,a[k];;ry);
         tadd(rx,ry;;z);       }
}

/* z += a[0]+a[1]+..+a[n-1] */
vsum(int n, int a[n]; int z; )
{ int z0=z; dcvsum(n,a;;r); tadd(z0,r;;z); }
```

Figure 19 A divide-and-conquer version of the routine `vsum` of Figure 17.

`vsum` of Figure 17. The routine `tadd` is that of Figure 12. The intermediate C code resulting from compiling the routine `dcvsum` is shown in Figure 20 and includes the three changes mentioned above.

As described elsewhere [Alternative][Dataflow][TSIA], the TSIA language allows for strict and non-strict evaluation. In this presentation, the TSIA language has a strict evaluation. As in most currently popular programming languages, each in argument of a routine is evaluated before the execution of the routine. With a strict evaluation, the `_ready` flag of a child task only depends on the code of its parent routine. A compiler thus can derive from the parent code an expression for the value of the `_ready` flag of each child task.

In a strict evaluation, the first child task of a routine is ready to execute. Thus `_pd1->_ready=1` in the compiled code for `dcvsum` in Figure 20. The second child task of a routine is ready to execute if it is independent of the first. For example, in the source code of `dcvsum` in Figure 19, the second task `dcvsum(n-k,a[k];;ry)` obviously is independent of the first task `dcvsum(k,a;;rx)`. Thus `_pd2->_ready=1` in the compiled code for `dcvsum` in Figure 20. In other words, the tasks `dcvsum(n-k,a[k];;ry)` and `dcvsum(k,a;;rx)` can execute in parallel.

```
typedef struct {void (*_f)(char **_psp); int _ready;
                int n, *a, *z; int _l;} _s_dcvsum;
void _f_dcvsum(char **_psp) {
  _s_dcvsum *_p = (_s_dcvsum *)*_psp;
  *_psp += sizeof(*_p);
  if (_p->n==1) *(_p->z) = *(_p->a);
  else {
    int _n=_p->n, *_a=_p->a, *_z=_p->z, k=_n/2;
    _s_tadd *_pa; _s_dcvsum *_pd2, *_pd1;

    *_psp -= sizeof(_s_tadd);
    _pa = (_s_tadd *)*_psp;
    _pa->_f = _f_tadd;
    _pa->_l = sizeof(_s_tadd);
    _pa->_ready = 0;
    /* _pa->z  = _z; */ /* Already true. */

    *_psp -= sizeof(_s_dcvsum);
    _pd2 = (_s_dcvsum *)*_psp;
    _pd2->_f = _f_dcvsum;
    _pd2->_l = sizeof(_s_dcvsum);
    _pd2->_ready = 1;
    _pd2->n  = k;
    _pd2->a  = _a;
    _pd2->z  = &(_pa->x);

    *_psp -= sizeof(_s_dcvsum);
    _pd1 = (_s_dcvsum *)*_psp;
    _pd1->_f = _f_dcvsum;
    _pd1->_l = sizeof(_s_dcvsum);
    _pd1->_ready = 1;
    _pd1->n  = _n-k;
    _pd1->a  = _a+k;
    _pd1->z  = &(_pa->y);
  }
}
```

Figure 20 Intermediate C code from compiling dcvsum of Figure 19.

Similarly, the third child task of a routine is ready to execute if it is independent of the first two. For example, in the source code of `dcvsum` in Figure 19, the third task `tadd(rx,ry;;z)` obviously is dependent on the second task `dcvsum(n-k,a[k];;ry)` and on the first task `dcvsum(k,a;;rx)`. Thus `_pa->_ready=0` in the compiled code for `dcvsum` in Figure 20. In other words, the task `tadd(rx,ry;;z)` cannot execute in parallel with the tasks `dcvsum(n-k,a[k];;ry)` and `dcvsum(k,a;;rx)`. Similarly, the expression for the `_ready` flag can be determined for the fourth and other child tasks of a routine.

The above description simplifies the compiler's derivation of an expression for `_ready`. For example, what if the second and third tasks of some routine are independent of each other, but each depends on the first task? If all three tasks are placed onto the stack, then a system task inserted between the first and second task could change `_ready` from `0` to `1` for the second and third tasks after the first task has executed. Alternatively, the parent could place just the first task onto the stack. Then after the execution of the first task, the resumed activation frame of the parent places the second and third tasks onto the stack, each with `_ready=1`.

In order to determine if two tasks are independent, all the effects of a task have to be declared [Dataflow][TSIA]. These include input/output, global items and other nonlocal effects. For example, a routine to print a character on standard output is declared as `putc(char c;;) (;stdout;)`. Nonlocal effects are declared in the second set of parentheses. Routines are declared in header files, as in C, or via other mechanisms. Then in the routine
   `putab(;;)(;stdout;) {putc('a';;);putc('b';;);}`
the `putc` tasks are not independent since each modifies `stdout`. The declaration of nonlocal effects also has other motivations [Scope].

The strict evaluation of this presentation requires the routine `putab` to declare the nonlocal effect `(;stdout;)`. Otherwise the example `putab(;;); putc('c';;)` could output the correct `abc` or the incorrect `cab`. In general, the declarations of nonlocal effects need to be propagated up the call chain of the application definition. Such propagation is tedious and error-prone if performed manually. Thus the propagation presumably would be performed automatically by the compiler or some other tool.

Instead of having the compiler transparently set the `_ready` flag, it could be explicitly set in the application definition. For example, this is essentially the purpose of the `spawn` and `sync` keywords of Cilk-2 through Cilk-5. These successors to Cilk-NOW support a parallel execution, but abandon transparency. Only a transparent execution is pursued in this presentation.

The `_ready` flag and the other two small changes to the compiled TSIA code allow for the following parallel execution.

Each computer processor of the parallel execution has its own stack, stack pointer and execution of the routine `_xtop`. The stacks are in shared memory.

The parallel execution of an application on three processors is shown in Figure 21. Each part a) through e) shows a snapshot of the three stacks.

The start of the application execution is illustrated in Figure 21a). One of the stacks contains the initial frame of the application and a few `_cop` frames. Each of the other stacks contains a `_thief` frame. In the example, the initial application frame is $d(9,b;;y)_1$, where for the convenience of the illustration, `d` is an abbreviation for the routine `dcvsum` of Figure 19. Similarly, later snapshots also use `t` an abbreviation for the routine `tadd`. The subscript in each frame illustrates the value of the `_ready` flag. In addition to or instead of the `_cop` frames, there could be other frames which 'clean up' once the application execution has completed.

When executed, a `_cop` frame will remove a corresponding `_thief` frame by replacing it with a `_skip` frame. As introduced in section 9, the execution of the `_skip` routine just removes its frame from the stack. In order to illustrate the correspondence, each `_cop` frame and `_thief` frame is labelled in Figure 21. For example, `_copA` corresponds to `_thiefA`. The correspondence is maintained by the `_cop` frame; it contains a pointer to the `_thief` frame.

When executed, a `_thief` frame steals the bottommost ready frame from a stack chosen at random. If there is no ready frame on that stack, the `_thief` tries another random randomly chosen stack. Since it is in an infinite loop, a `_thief` never removes itself from the stack. A `_thief` frame only can be removed from the stack by its corresponding `_cop`.

The execution of the initial application frame $d(9,b;;y)_1$ in Figure 21a) yields the frames $d(4,b[0];;r03)_1$, $d(5,b[4];;r48)_1$, $t(r03,r48;;y)_0$, as shown in Figure 21b). The illustration emphasizes the new or changed frames between the snapshots.

As for any arc in the dag of an application execution, the name of the item `r03` is arbitrary in Figure 21b). It is simply a convenient means to refer to a location in the frame `t(r03,r48;;y)`. The name `r03` is used in the illustration since it corresponds to the elements `b[0]` through `b[3]` in the task d(4,b[0];;r03). Similarly, the names of other items in Figure 21 are arbitrary and are chosen for the convenience of the illustration.

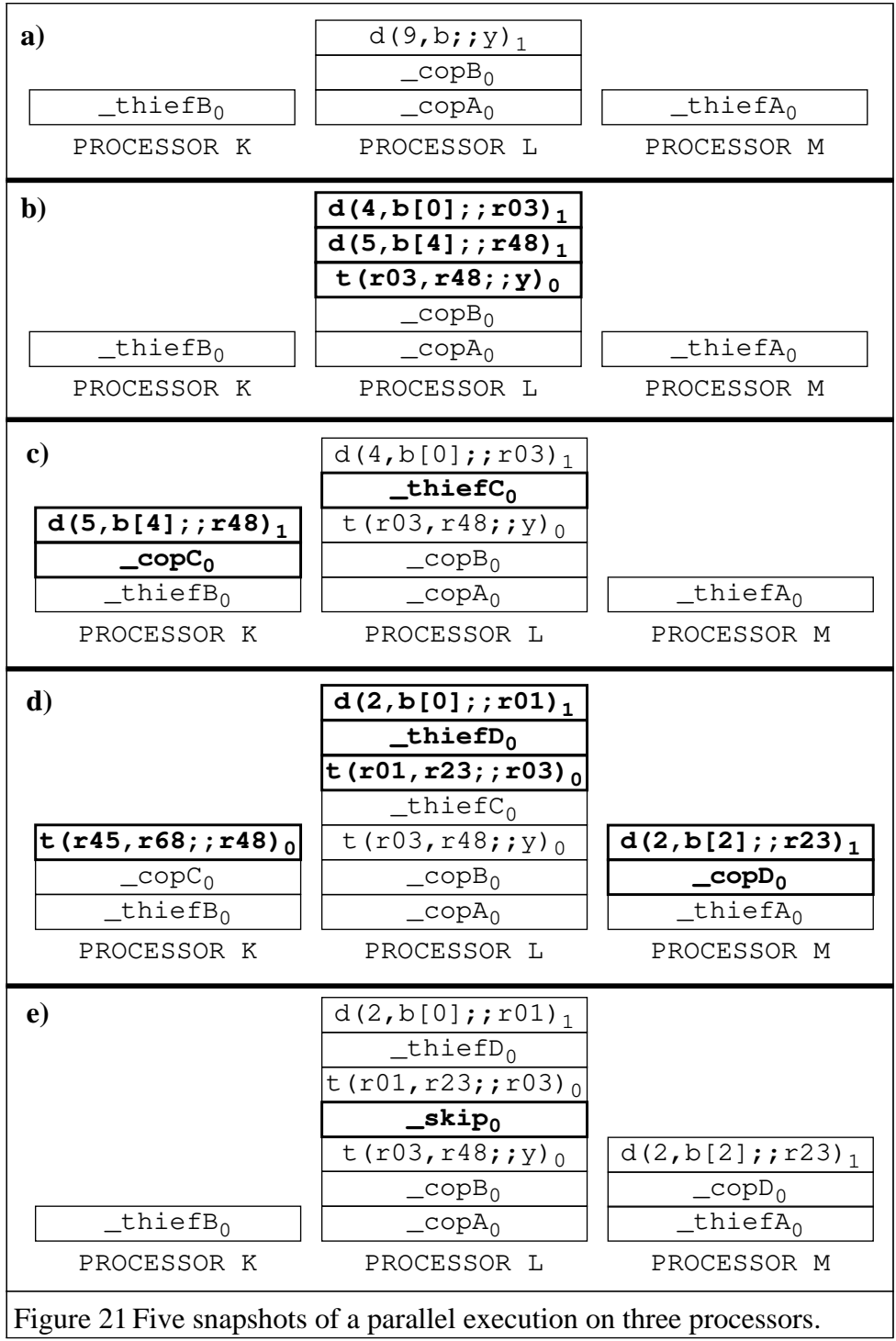

Figure 21 Five snapshots of a parallel execution on three processors.

The snapshot in Figure 21c) assumes an execution where _thiefB has successfully stolen the frame d(5,b[4];;r48)$_1$ from the stack of PROCESSOR L. _thiefB chose to steal that particular frame since it was the bottommost ready frame on that stack. After finding the frame, _thiefB moved the frame to its own stack. On the stack of PROCESSOR L, _thiefB replaced the stolen frame by the frame _thiefC. The length of the _thiefC frame thus is that of the stolen frame. The corresponding _copC frame was placed by _thiefB on its own stack, below the stolen frame.

By the definition of a ready frame, no task writes into a ready frame. Thus no task has the address of any item of a ready frame. Thus a ready frame and its items can be moved from one area of memory to another. In other words, the definition of a ready frame allows _thiefB to move the ready frame d(5,b[4];;r48)$_1$ to its own stack. By contrast, if the not-ready frame t(r03,r48;;y)$_0$ were moved in memory, then the out r48 of the task d(5,b[4];;r48)$_1$ would be invalid.

The _thiefC frame serves as a barrier. Below _thiefC, the frames require the results of the stolen task d(5,b[4];;r48)$_1$. In particular, the out r48 is required by the frame t(r03,r48;;y)$_0$. Since the results of the stolen task are not yet available, _thiefC ensures that the frames below it are not yet executed. If ever reached in the execution, _thiefC tries to steal a ready frame, so that PROCESSOR L is not idly waiting for the results of the stolen task.

When the execution of the stolen frame d(5,b[4];;r48)$_1$ and all its descendents has completed, then its results are available. In particular, the out r48 is written to the frame t(r03,r48;;y)$_0$. Since the results of the stolen task are now available, _thiefC must be removed in order to allow the frames below _thiefC to executed. Thus, when the stolen frame d(5,b[4];;r48)$_1$ and all its descendents have executed, the frame _copC is executed to remove the frame _thiefC.

In the snapshot in Figure 21d), the frame t(r45,r68;;r48)$_0$ is all that remains on the stack of PROCESSOR K from the execution of the stolen frame d(5,b[4];;r48)$_1$ and its descendents.

Also in Figure 21d), the frame d(4,b[0];;r03)$_1$ of Figure 21c) has executed on PROCESSOR L, yielding the frames d(2,b[0];; r01)$_1$, d(2,b[2];;r23)$_1$, t(r01,r23;;r03)$_0$. Subsequently,

Figure 21d) assumed an execution where `_thiefA` of `PROCESSOR M` successfully stole the frame `d(2,b[2];;r23)`$_1$ from the stack of `PROCESSOR L`.

In the snapshot in Figure 21e), the frames `t(r45,r68;;r48)`$_0$ and `_copC` of Figure 21d) have executed on `PROCESSOR K`. Thus on the stack of `PROCESSOR L`, `_thiefC` has been replaced by `_skip`.

Though not further illustrated in Figure 21, the execution would continue in a similar fashion, until the only application frame left would be `t(r03,r48;;y)`$_0$ on the stack of `PROCESSOR L`. After its execution, `_copB` and `_copA` execute. With no frames left on any stack, each of the `_xtop` executions exits and control is returned to the user of the application.

The above outline does not describe some necessary protocols. One protocol ensures that at most one `_thief` operates on any given stack.

Another protocol is between a `_thief` and the `_xtop` of a stack. Though they start at opposite ends of the stack--`_xtop` at the top--`_thief` at the bottom--`_xtop` and `_thief` can collide if there are few or no ready frames on the stack. Since each of their actions can modify the stack, `_xtop` and `_thief` have to obey a protocol in order to survive possible collisions. Since `_xtop` executes every task of the application, it is very important that the protocol adds minimal overhead to `_xtop`. An example of such a protocol is described elsewhere [Cilk-5].

By contrast, no protocol is needed for a `_cop`; it can simply write the address of the `_skip` instruction into the `_thief` task frame. No collisions are possible since a `_thief` never modifies its own frame. All that is required is for the `_thief` to occasionally return to `_xtop` and/or poll the instruction in its own frame.

For a sequential application execution on a single processor, there are no `_thief` frames. The execution essentially is that described in the previous sections. The differences are the three small changes to the compiled TSIA code described at the beginning of this section and the additional protocol required by `_xtop` mentioned above. The differences introduce little overhead to the application execution. As desired by a transparent parallel execution, TPARX efficiently executes an application on a single processor.

For a parallel application execution on multiple processors, the `_cop` and `_thief` frames implement work-stealing scheduling. As described in the next section, the work-stealing scheduling of TPARX is very similar to that of Cilk-NOW and its precursors. That scheduling is space, time and communication efficient in theory and in practice [Cilk-1]. As desired by a transparent parallel execution, TPARX thus is expected to efficiently execute an application on multiple processors.

The efficiency is largely a result of the depth-first execution on each processor and the breadth-first work-stealing [Cilk-1]. Depth-first corresponds to the topmost task frame and is the sequential execution order. Breadth-first corresponds to the bottommost task frame and tends to steal work large in quantity and along the critical path.

As promised, the parallel execution implemented by TPARX is transparent. The application source code of Figure 19 contains nothing concerning parallelism. Instead, all the details concerning parallelism are in TPARX, a system external to the application.

TPARX largely consists of the three small changes to the compiled TSIA code described at the beginning of this section as well as the routines `_cop`, `_thief`, `_skip` and `_xtop`. TPARX thus is a very small and simple system. TPARX is an example of an external system well served by the structured application execution offered by task frames.

The structured execution is maintained by TPARX. As illustrated by the snapshots in Figure 21, the parallel execution moves from one structured state to another. The structured execution allows parallelism to be easily combined with other execution features. For example, the structured parallel execution easily allows for a debugger for the application definition. Similarly, the parallel execution easily is extended to an adaptive execution, where the number of processors available for the application execution varies during the course of the execution.

TPARX is suitable as is for a variety of applications. Many of these applications use divide-and-conquer algorithms. The applications include sorting and dense matrix algorithms [Cilk-5][TSIA]. This variety of applications is promising since TPARX is just an initial implementation of a system providing a transparent parallel execution.

Compared to the above initial success, the ultimate success of TPARX will depend on how well it can be extended to support other applications. The support of an application implies a convenient application definition and an efficient application execution. Two examples requiring extensions to TPARX follow. As is, TPARX allows an application like search to speculatively attempt many possible solutions in parallel, but it does not allow unnecessary attempts to be aborted once a solution is found. Such applications are described elsewhere [Cilk-5], as is their convenient definition for a transparent execution [TSIA]. As is, the locality of TPARX's scheduling is good enough for many applications on a symmetric multiprocessor
(SMP) [Cilk-5]. However, on processors with distributed memory, the
locality may not be good enough compared to an execution which explicitly places data in the memory of particular processors. The possibility of
such placement in a transparent execution with a convenient application
definition is described elsewhere [Dataflow][TSIA].

## 11 Some Related Work

The alternative execution of routines proposed in this presentation is a
result of TSIA, a model for transparent application execution. TSIA covers
a large area of computing. References to some work in the area can be
found elsewhere [Alternative][Dataflow][Dividing][TSIA]. Even just the
alternative execution of routines covers a large area of computing. A survey
of this area is beyond the purpose of this proposal. Instead, this section
briefly compares the alternative execution of routines to some related work.

In the alternative execution, routines execute using a stack of task
frames. In Cilk-NOW and its precursors, routines execute using task
frames in a structure similar to a stack [Cilk-1][Cilk-NOW][PCM]. Cilk-NOW and its precursors arguably are the most closely related work to that
of this presentation. Like TPARX of the previous section, Cilk-NOW and
its precursors provide an application with a transparent parallel execution.

A task is defined by its nonblocking execution. In many real-world systems and in many research systems, a simple application executes in terms
of tasks [Dividing][TSIA]. However, Cilk-NOW and its precursors allow
various applications to execute in terms of tasks. This variety arises from
the support for an executing task to create task frames. In other words, in its
execution a task frame can replace itself by other task frames. Without this
mechanism, a task cannot call a routine as a task. Without calls to routines,
only simple applications can execute in terms of tasks. The mechanism is
known as delegation in other presentations of TSIA. Delegation also offers
other benefits [Alternative].

Delegation is a variation on continuation, a technique from the Scheme
programming language and other functional computing [RABBIT]. For
example, each allows proper tail calling. A continuation is based on control-flow. A delegation is based on dataflow. In other words, the dependencies between routines are implicit for a continuation, while for a delegation
the dependencies are explicit. The mechanism of delegation also is used in
some implementations of graph reduction [ALICE].

In Cilk-NOW and its precursors, task frames are kept in a heap, but each processor maintains a queue of ready frames. Since it preserves the application's hierarchy of calls to routines, the queue is similar to the stack of this presentation. Each processor performs a depth-first execution of the frames on its local queue. Frames are stolen breadth-first from a remote queue.

In Cilk-NOW and its precursors, a task frame does not have a `_ready` flag like that of TPARX of the previous section. Instead, each frame has a join counter indicating the number of missing arguments needed for the frame to be ready to execute. Thus whenever an executing task produces an out, it decrements the join counter of the frame using that out as an in. If the join counter goes to zero, the frame is added to the queue. The join counter is more general than the `_ready` flag, but few Cilk applications use this extra generality. In fact, the extra generality is assumed forbidden in the proofs of Cilk's efficient scheduling. The proofs require an application to have a strict execution. The join counter is costlier than the `_ready` flag. The join counter requires considerable effort at run-time. The `_ready` flag generally can be determined at compile-time.

The work-stealing scheduling of TPARX is similar to, but not the same as, that of Cilk-NOW and its precursors. For example, when an executing task reduces a join counter to zero, the resulting ready frame is recognized and is posted on a queue of ready frames. By contrast, in TPARX such a ready frame is not recognized. As is, TPARX thus does not meet an assumption of Cilk's proofs of efficient scheduling. The practical effect of these differences in scheduling can be determined once TPARX and applications are implemented. If necessary, a more sophisticated `_cop/_thief` synchronization can recognize such ready frames; each is effectively a topmost task on its stack.

## 12 Summary

The alternative execution of routines places task frames onto a stack. A task frame is the call of a routine to be executed. By contrast, the current conventional execution of routines places activation frames onto a stack. An activation frame is the internal state of an executing routine.

A crude implementation of the alternative execution demonstrates the feasibility of the alternative execution, including the following two aspects. The current conventional execution of routines can be treated as a special case of the alternative execution. For many applications, the alternative execution of routines does not introduce significant time overheads beyond those of the current conventional execution.

As outlined, the crude implementation can be extended to a system called TPARX which provides a variety of applications with a transparent parallel execution. TPARX is a very small and simple system. As demonstrated by TPARX, the alternative execution of routines allows an external system to provide an application with a transparent execution.